\documentstyle[aaspp4]{article}

\lefthead{Elmegreen, Kim, and Staveley-Smith}
\righthead{A Fractal Analysis of the HI Emission from the Large
Magellanic Cloud}

\slugcomment{{\it Astrophysical Journal}, 548, Feb 10, 2001}

\begin{document}

\title{A Fractal Analysis of the HI Emission from the Large
Magellanic Cloud}

\author{Bruce G. Elmegreen\altaffilmark{1},
Sungeun Kim\altaffilmark{2,3},
Lister Staveley-Smith\altaffilmark{4}}

\altaffiltext{1}{IBM Research Division, T.J. Watson Research Center, P.O.
Box 218, Yorktown Heights, NY 10598; e--mail: bge@watson.ibm.com}
\altaffiltext{2}{Astronomy Department, University of Illinois at Urbana-
Champaign, IL 61801}
\altaffiltext{3}{Harvard Smithsonian Center for Astrophysics, 60 Garden Street,
Cambridge, MA 02138; email:skim@cfa.harvard.edu} 
\altaffiltext{4}{Australia Telescope National Facility, CSIRO, PO Box 76,
Epping, NSW 1710, Australia; email lstavele@atnf.csiro.au}

\begin{abstract} Fourier transform power spectra of the distribution of
neutral hydrogen emission in the Large Magellanic Cloud is approximately a
power law over $\sim2$ decades in length. Power spectra in the azimuthal
direction look about
the same as the rectilinear spectra. No difference is seen between power
spectra of single channel maps and power spectra of either the peak
emission map or the integrated emission map at the same location. 
There is a slight steeping of the average 1D and 2D LMC power spectra at high
spatial frequencies. Delta-variance methods also show the same power-law structure.

These results suggest that most of the interstellar medium in the LMC is
fractal, presumably the result of pervasive turbulence, self-gravity, and
self-similar stirring. The similarity
between the channel and integrated maps suggests they cover about the
same line-of-sight depth. The slight steepening of the power spectra
at high spatial frequency, corresponding to wavelengths smaller than
$\sim100$ pc, could mark the transition from large-scale emission that is
relatively shallow on the line of sight to small-scale emission that is
relatively thick on the line of sight. Such a transition, if real, would
provide a method to obtain the thickness of a face-on galactic gas layer.

To check this possibility, three dimensional fractal models are made
from the inverse Fourier transform of noise with a power-law cutoff.
The models are viewed in projection with a Gaussian density distribution
on the line of sight to represent a face-on galaxy disk with finite
disk thickness. The density structure from turbulence is simulated in
the models by using a log-normal density distribution function with a
scale factor dependent on the Mach number.  Additional density structure
from simulated HI phase transitions is included in some models.
After tuning the Mach number, galaxy thickness, and mathematical form of
the phase transition, the models can be made to reproduce the observed
LMC power spectra, the amplitude of the HI brightness fluctuations,
and the probability distribution function for brightness.  In all cases,
the HI structure arises from a relatively thin layer in the LMC; the thick
part of the HI disk has little spatial structure.  The large amplitude
of the observed intensity variations cannot be achieved by turbulence
alone; phase transitions are required.

The character of the fractal HI structure in the LMC is viewed
in another way
by comparing positive and negative images of the integrated emission.
For the isotropic fractal models, these two images have the same general
appearance, but for the LMC they differ markedly. The HI is much more
filamentary in the LMC than in an isotropic fractal, making the geometric
structure of the high-emission regions qualitatively different than the
geometric structure of the low-emission (intercloud) regions.  The 
high-emission regions are also more sharply peaked than the low-emission
regions, suggesting that compressive events formed the high-emission
regions, and expansion events, whether from explosions or turbulence,
formed the low-emission regions.

The character of the structure is also investigated as a function of scale
using unsharp masks and enlargements with four different resolutions.
The circular quality of the low emission regions and the filamentary quality
of the high emission regions is preserved on scales ranging from several
tens to several hundreds of parsecs.

The spatial scales for sources of turbulent energy input may be illustrated
by rms variations in the power spectra with position in the galaxy.  This rms
decreases from $\sim0.6$ at kpc scales to $\sim0.25$ on $\sim20$ pc scales.
The large scale variations are probably from known supershells. The
smaller scale variations could be the result of a combination of turbulent
cascades from these large-scale energy inputs and additional energy
sources with smaller sizes.  \end{abstract}

Subject Headings:  turbulence -- ISM: clouds -- ISM: structure -- 
galaxies: ISM -- Magellanic Clouds

\section{Introduction} 

Power spectra of HI in the Small Magellanic Cloud show a power
law decrease with increasing wavenumber over 2.3 decades in length
(Stanimirovic et al. 1999). Box-counting techniques and perimeter-area
measures of the HI in M81 group galaxies indicate a scale-free behavior
there too, over 1.7 decades (Westpfahl et al. 1999).  These results
for whole galaxies extend to much larger scales the fractal nature of
interstellar gas that was previously found in Milky Way clouds using the
same methods of power spectra (Crovisier \& Dickey 1983; Green 1993;
Lazarian \& Pogosyan 2000; Stutzki et al. 1998) and perimeter-area
correlations (Falgarone, Phillips, \& Walker 1991) . The similarity of
these gas structures to the distribution of smoke particles and other
passive tracers in incompressible laboratory turbulence (Sreenivasan
1991) and to the distribution of density in simulations of compressible
turbulence (Elmegreen 1999; Rosolowsky et al. 1999; MacLow \& Ossenkopf
2000; Pichardo et al. 2000) imply that turbulence is an important
structuring agent for gas on a wide range of scales in a galaxy disk.
Most of the galactic gas could be fractal as a result of this
turbulence, producing both clouds with their power-law mass
spectra (Elmegreen \& Falgarone 1996; Stutzki et al. 1998)
and the low density intercloud medium (Elmegreen 1997).

This paper applies a power spectrum analysis to HI emission from the
Large Magellanic Cloud using the combined Australia Telescope Compact
Array (ATCA) interferometric survey of HI by Kim et al. (1998a,b, 1999)
and a single dish survey done with the Parkes Multibeam by 
Staveley-Smith et al. (in preparation). The combined data set is 
described by Kim et al. (2000).
Because the HI is more structured in the LMC
than the SMC as a result of supershells (Meaburn 1980, Kim et al. 1999)
and other effects of star formation,
we examine
both local and global power spectra from one dimensional cuts in right
ascension and declination, as well as power spectra in the azimuthal
direction. We also determine the more conventional 2D power spectrum
and Delta-variance for the whole LMC. 
The results are compared with power spectra in 
the same field from foreground Milky Way gas. 
A model fractal medium is used
to comment on the implications of the observations.

One of the results of this work is that the structure of HI in the
LMC is not a simple fractal like that obtained from fractal Brownian
motion models (e.g., Stutzki et al. 1998).  The geometry of the high
temperature emission is distinctly different from the geometry of the
low temperature regions.  Moreover, the extremes in HI brightness are
too large to be reproduced by a realistic isothermal turbulence model. Rather, the
ISM looks like a self-similar superposition of shells and filaments,
separated by roundish holes. The contrast between the emission
regions and the holes seems to require at least two phases of gas,
such as the warm and cool phases of HI.  In this respect, the LMC 
resembles computer simulations by Wada, Spaans, \& Kim (2000).  These
simulations reproduce the large shells and the amorphous structures that
are between the shells.  Similar models and a more detailed study of
the energy input from overlapping shells were made by Scalo \& Chappell
(1999a,b).

\section{Data Source}

The ATCA was used in the mosaic mode to survey a region 10$^{\circ}$
$\times$ 12$^{\circ}$ covering the LMC, and at an angular resolution
of $1^{\prime}$, corresponding to a spatial resolution of 15 pc (for
an assumed distance of 50 kpc). The detailed observations and data
reduction are described in Kim et al. (1998a,b). The ATCA data for the LMC
have been combined with Parkes single-dish data in order to
add the low spatial frequencies missed by the interferometer
(Kim, Staveley-Smith \& Sault 2000).  The Parkes observations were taken
with the inner 7 beams of the Parkes Multibeam receiver
(Staveley-Smith et al. 1996) in December 1998. The telescope has a
half-power beamwidth of 14\farcm1. A total bandwidth of 8 MHz was used
with 2048 spectral channels in each of two orthogonal linear
polarizations. The velocity spacing of the multibeam data is 0.82 km
s$^{-1}$, but the final cube was Hanning-smoothed to a resolution of
1.6 km s$^{-1}$.  The velocity range in the trimmed final cube is $-66$ to 430
km s$^{-1}$ with respect to the barycentric reference frame.
The pixel size in the final maps used here is $40^{\prime\prime}$, 
corresponding to $\sim10$ pc. 

The shortest baseline (30 m) of the ATCA and the diameter (64 m) of
the Parkes telescope provides an excellent overlap in the Fourier domain,
allowing adequate cross-calibration of data from the two sources. 
The calibration agrees to within a few percent of the prediction (Kim et al.
in preparation).

\section{Analysis} 

The HI emission maps of the LMC show a complex structure composed
of holes, filaments, and clumps, all with a wide range of scales.
A previous analysis by Kim et al. (1999) of the holes and shells in the HI
distribution found evidence for expansion around many of these sources,
with a correlation between the expansion speed and the size of the
shell that was consistent with energy input from visible massive stars.
However, the majority of emission regions are not so clearly identified
with energy sources.  There is a lot of amorphous structure between
the shells, some of which is filamentary, and much of this could be
caused by random compressive events related to supersonic turbulence.

Turbulence produces a fractal-like pattern (Mandelbrot 1983)
in which large regions are
broken up into smaller regions, and these are broken further into even
smaller regions, producing an overall hierarchy of structures.  It is
not possible to isolate individual ``clouds'' in such a structure without
introducing some type of bias about what a cloud is and about how nested
clouds should be counted.  A better way to analyze this emission is with
Fourier transform power spectra (PS), defined as the sum of the squares of
the real and imaginary parts of the Fourier transform of the emission map,
plotted versus the wavenumber, $k=2\pi/\lambda$ for wavelength $\lambda$
of certain features.  Power spectra indicate the relative importance of
features of various sizes. If a structure has no dominant scale, as would
be the case for a perfect fractal, then the PS would be a power law
without kinks or features either.  For an emission region of this type,
the structure cannot be characterized as consisting of discrete and
isolated clouds with certain sizes, but should be viewed as amorphous
with a measurable parameter equal to the slope of the PS.

Fourier transform power spectra have been used to analyze HI structure
in the local Galaxy by Crovisier \& Dickey (1983) and Green (1993).
They found power-law behavior indicative of a scale-free emission
distribution.  Analogous techniques have been applied to molecular data
by Stutzki et al. (1998), who also got a scale free result.  The first
power spectrum study of a whole galaxy was by Stanimirovic et al. (1999)
who used the ATCA + Parkes data of the Small Magellanic Cloud.  Here we
perform similar analyses of the LMC data from this same
telescope combination.

In what follows, the 
power spectra for the LMC 
and foreground Galactic emission 
are presented in Section
\ref{sect:ps}. The implications for the line-of-sight distribution and
depth of the HI, and for the scale of turbulent energy sources, are discussed
in Sections \ref{sect:ch} - \ref{sect:en}.  Power spectra in the azimuthal
direction are presented in Section \ref{sect:az},
and two-dimensional power spectra, along with the two-dimensional Delta-variance,
are in Section \ref{sect:2dps}. 

Power spectrum and Delta-variance analyses of spatially limited maps
can introduce errors if there are sharp edges, but when the emission
tapers off slowly toward the edges, as it does for whole galaxies like
the LMC, then these problems are minimal.  Power spectra
of azimuthal profiles have no edge effects either.  For this reason, our
main results are not affected by map-edge effects. The only discussions
where these effects might be important are those involving the $3\times3$
segments of the whole galaxy, which are used to determine rms variations
with position, and those involving Milky Way emission on the line
of sight to the LMC.

Fourier transform power spectra have other limitations too.  Section
\ref{sect:fi} shows the LMC data in a different way to illustrate
something that does not appear in the power spectra, namely the
different topologies of the cloud and intercloud regions.  The cloudy
or high-brightness regions tend to be filamentary, while the intercloud
or low-brightness regions tend to be round.  Section \ref{sect:ga} also
shows another feature of the self-similar geometry using unsharp masks
with four different scales of smoothing.

\subsection{Power Spectra}
\label{sect:ps}

Fourier transform power spectra (PS) were made of the channel maps, peak
temperature map, and velocity-integrated map of HI emission from the Large
Magellanic Cloud. Figure \ref{fig:images} 
shows the corresponding emission
distributions. Each panel contains a field comprised of 600 pixels in
the North-South direction by 630 pixels in the East-West direction,
in addition to subfields measuring $200\times210$ pixels, indicated by white
lines. The pixel size in the figures is $40^{\prime\prime}$; 
note that the beam size is $60^{\prime\prime}$
and the original sampling interval was $20^{\prime\prime}$. 
The map in the top left shows the peak temperature, the top
right shows the velocity-integration, the lower left is the $v=254\pm 0.8$
km s$^{-1}$ channel and the lower right is the $302\pm 0.8$ km s$^{-1}$
channel. Nine other channel maps were examined too, with 16 km s$^{-1}$
spacing, and found to give similar results to the 254 and 302 km s$^{-1}$
maps. 
The long stream of emission south of 30 Doradus is brighter
on the integrated map than the peak temperature map because of its larger
linewidth compared to the rest of the galaxy; it may contain
a contribution from material that is not in the disk. 

These images are optimized for the display of features and are not calibrated
with a linear conversion between intensity and darkness. The conversion
is close to linear, but the faint gray scales are highlighted a bit. 
For calibrated HI maps, see Kim et al. (2000).

\begin{figure}
\vspace{6.in}
\includegraphics{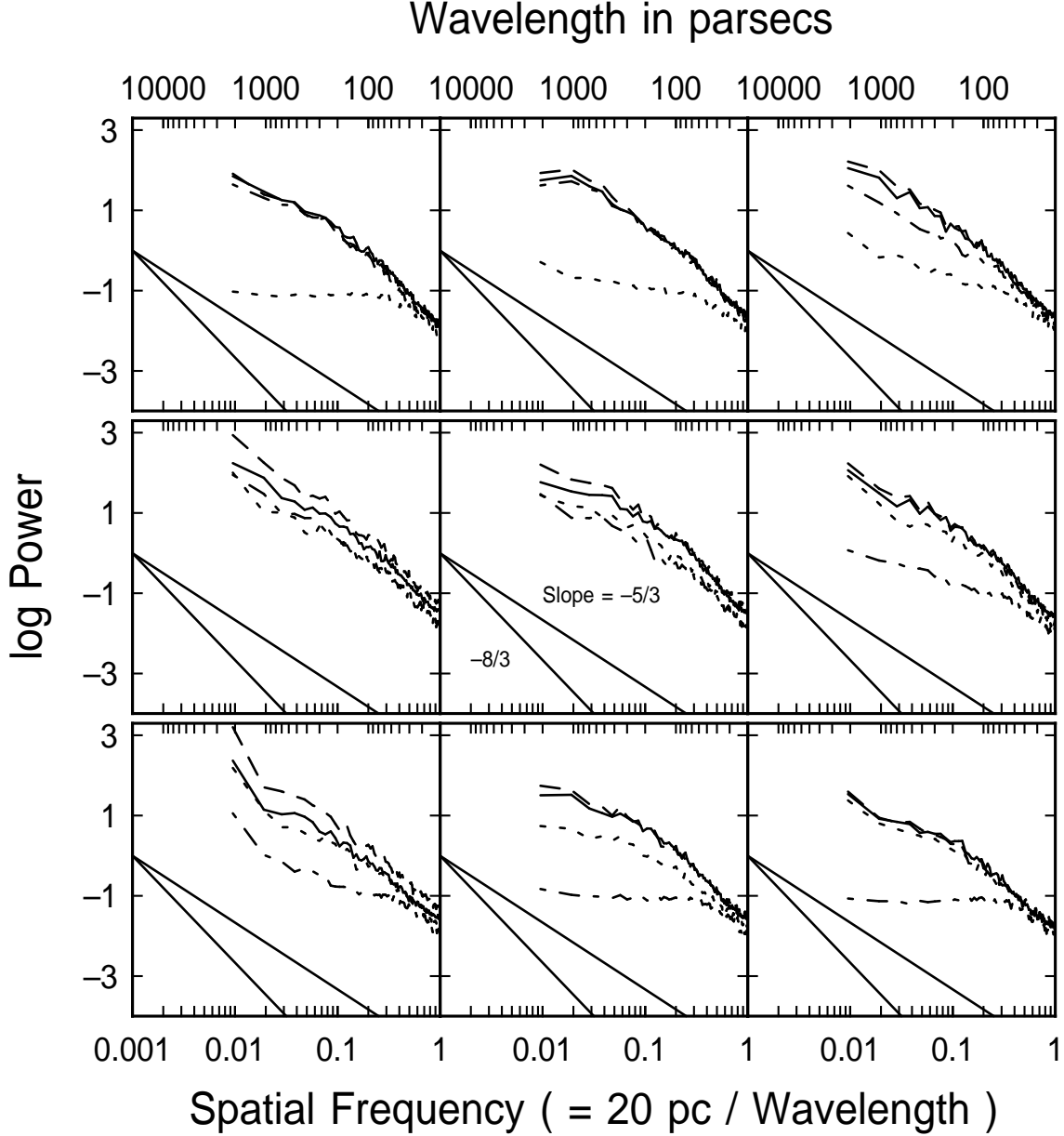}
\caption{Average,
one-dimensional power spectra of the HI emission from each of the $3\times3$
panels in figure 1, shown here in a similar $3\times3$ array with the
same orientation (North up). 
Lines types denote the various HI maps shown in Figure 1:
solid = peak temperature map, dashed = velocity integrated map, 
dotted = 254 km s$^{-1}$ channel map, and dot-dashed = 302 km s$^{-1}$ channel map.  
Kolmogrov turbulence is expected to
give a 1D Power spectrum with a slope of $-5/3$ if the region observed
is thinner on the line-of-sight than the transverse wavelength.  This is
the slope seen here at small wavenumbers. The same turbulence is expected to
give a 1D Power spectrum with a steeper slope of $-8/3$ if the region
is thicker on the line-of-sight than the wavelength, and this is what
the figure suggests for wavenumbers exceeding 0.2. The dividing point
corresponds to a wavelength of $\sim100$ pc, which is suggested to be
a measure of the line-of-sight thickness of the HI layer in the LMC.}
\label{fig:ps}
\end{figure}

\begin{figure}
\vspace{6.in}
\includegraphics{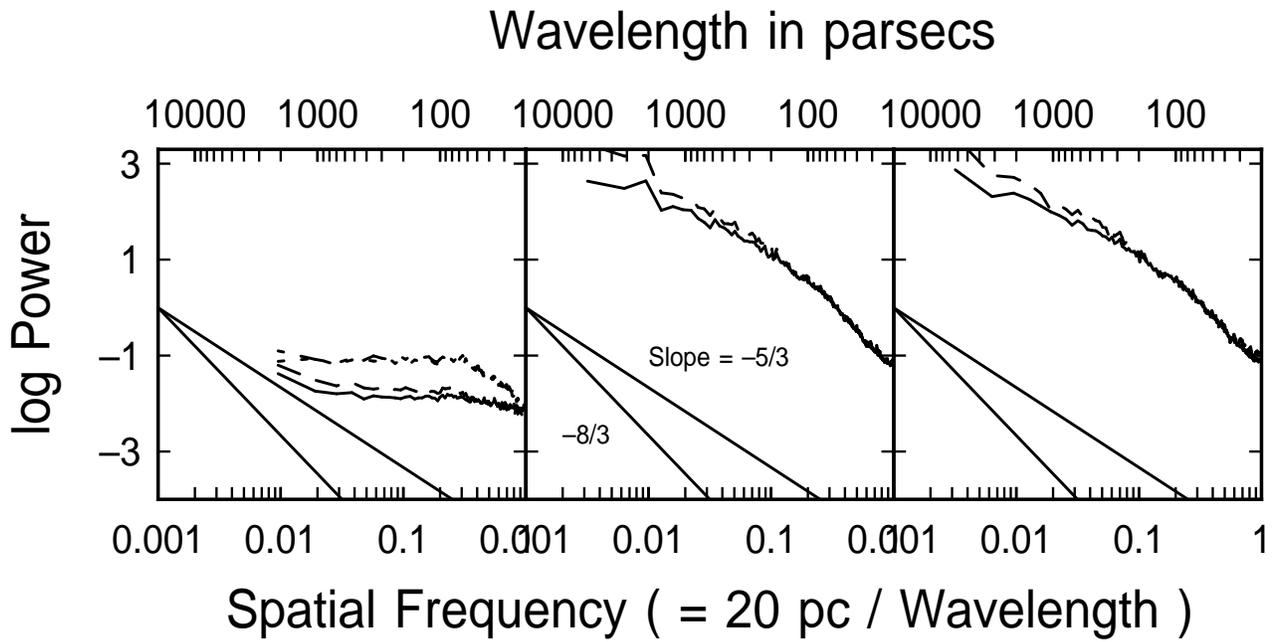}
\caption{Left: Average,
one-dimensional power spectra of the HI emission from 
the sky region shown in 
figure 4; middle: average 1D power spectrum for the whole galaxy,
using 1D FFTs in the horizontal (E-W) 
direction, and, right:  power spectrum of the whole galaxy using 1D
FFTs in the vertical (N-S) direction.
Line types are as in Fig. 2. 
}
\label{fig:psb}
\end{figure}

Figures \ref{fig:ps} 
and \ref{fig:psb}
show 1D power spectra for the LMC maps. Each PS is the
average of many one-dimensional PS taken for the horizontal rows of pixels
in the corresponding field. Solid lines are for the peak temperature
data, dashed lines are for the integrated data, dot-dashed lines are
for the 302 km s$^{-1}$ channel and dotted lines are for the 254
km s$^{-1}$ channel. For the 3$\times$3 subfields of the galaxy shown in
the corresponding 3$\times$3 panels of Figure \ref{fig:ps},
the PS of each 210-pixel row was obtained from the FFT of the intensity
using the IBM ESSL subroutine library, and then the PS from all 200 rows
were averaged together to give the plotted result.

The same $210\times200$ field size was used for a sky region northeast of
the galaxy.  The sky region is shown on the integrated map in Figure \ref{fig:sky}, 
and the sky PS is shown in the left panel of Figure \ref{fig:psb}.
This sky PS indicates the noise level for most of our study. At all but the
highest frequencies, the noise contributes only a percent or less to the
PS signal. 

The middle panel in Figure \ref{fig:psb} 
shows the average of the 1D
power spectra for the whole galaxy field, made by averaging 600 power
spectra from the 630-pixel-long FFTs of the horizontal rows. The
right panel shows the average 1D power spectrum taken in a vertical
direction, i.e., using 600 FFTs of 630-pixel long columns in the
north-south direction (and using a different LMC boundary than what
Fig. \ref{fig:images} 
shows, i.e., $600\times630$ pixels compared to
$630\times600$ pixels).

The absolute brightnesses of the channel, peak temperature, and
integrated maps are all different, so Figures \ref{fig:ps} 
and \ref{fig:psb}
normalize
the PS to lie at approximately the same height for easy comparison.
The greater linewidth of the field south of 30 Dor gives it a larger
ratio of integrated emission to peak temperature than the other fields,
so the PS of the integrated emission south of 30 Dor shifts upward
compared to the PS of the peak emission in this region.

The wavenumber ranges for all of the power spectra in figures \ref{fig:ps}
and \ref{fig:psb}
are also normalized so that image variations on a pixel scale are on the
right-hand side of each abscissa at a normalized spatial frequency of
1, while variations on larger scales are on the left of each abscissa,
at fractional normalized wavenumbers. The wavelength corresponding to a
particular spatial frequency $k$ equals 2/k pixels, which is about $20/k$
parsecs, as shown on the top axes. Thus each of the 3$\times$3
panels in Figure \ref{fig:ps}
have PS that extend from a 20 pc scale on the right of the abscissa
to a $105\times20=2100$ pc scale near the left of the abscissa. 
The 
middle and right panels of Figure \ref{fig:psb}
go to $3\times$ larger spatial scales, so the
PS extend further to the left in the plots.

Power spectra of foreground HI emission is shown in Figure
\ref{fig:foreground}.  The three solid lines are the power spectra for
HI emission in $200\times220$ pixel fields on the same line of sight as
the LMC but at different velocities, covering foreground emission from
the Milky Way. The dashed line is the PS from the peak temperature map
of the LMC. The Milky Way emission has been divided by 40 to place the
PS in about the same range as the peak temperature PS.  The foreground
emission has a PS that curls up at large spatial frequency, presumably
from noise.  The Milky Way emission has a generally steeper slope as well.
Other correlation studies of high-latitude Galactic gas based on IRAS observations
are in Abergel et al. (1996).

\subsection{Channel Maps the same as Integrated Maps}
\label{sect:ch}

Figures \ref{fig:ps} and \ref{fig:psb}
contain several interesting results. First, the
power spectra of the channel maps are essentially the same as the power
spectra of the peak temperature and integrated maps in the directions
where the channel emission originates. For example, most of the emission
from the Northeast subfield is near 302 km s$^{-1}$, so the PS of the peak
or integrated maps are about the same as the PS of the channel map in
the top left of the 3$\times$3 panels. Similarly, most of the emission
in the Southwest is near 254 km s$^{-1}$, so the PS of that channel is
similar to the PS of the peak temperature or integrated maps in the
lower right of the 3$\times$3 panels.

This agreement between channel PS and peak-T or integrated PS occurs
even though the channels are only a small part of the total emission.
The implication of this result is that essentially all of the structure
on a wide range of scales contributes equally to the channel maps at
the velocity representative of the centroid speed of that region. The
high velocity tail of the whole line spectrum does not, for example,
contain only the small scale features. If it did, then the channel at
the centroid would have relatively stronger large scale structure than
the integrated map.

A related result is that the channel map cannot represent any thinner
emission on the line of sight than the peak-T or integrated maps. Here,
thinner is in the sense defined by Lazarian \& Pogosyan (2000) as
representing a spatial extent on the line of sight that is much smaller
than the wavelength of the PS in the plane of the sky. If a channel map
were thin in this sense, and the peak-T and integrated maps were not,
then the PS of the channel map would be flatter than the PS of the peak-T
or integrated maps (Lazarian \& Pogosyan 2000). Since the PS are all
about the same, the channel maps have emission from about the same 
line- of-sight depth as the integrated map.

The same is true for the peak-T and integrated maps: the similarity of
their power spectra over all spatial frequencies means that they are
either both shallow or both thick on the line of sight.  For example,
the peak temperature emission does not come from a thin midplane and
the integrated emission from a thick disk.

\subsection{Steepening at High Spatial Frequency as Evidence for
Finite Disk Thickness}
\label{sect:th}

\begin{figure}
\vspace{3.in}
\includegraphics{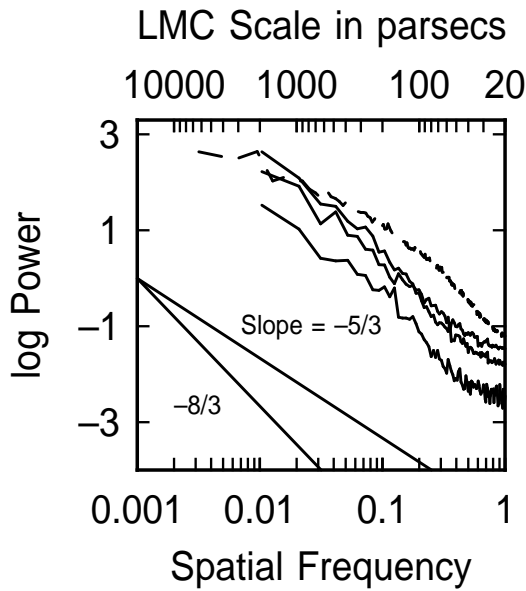}
\caption{Power spectra of foreground HI emission from the Milky Way
in three different velocity channels.  The HI power spectrum is
generally steeper for foreground emission than it is for the low
wavenumber part of the LMC emission, but it is similar to the
high-wavenumber part of the LMC, suggesting that the foreground
emission in each 
velocity channel is physically thick on the line of sight, like the LMC at
short wavelengths.}
\label{fig:foreground}
\end{figure}

The power spectra of the LMC are approximately power laws in two separate
segments, with a larger slope at larger spectral frequencies.  This is
not the case for the Milky Way power spectra, which have a constant
slope (cf. Fig. \ref{fig:foreground}).  The curvature with increasing
slope in the LMC PS is consistent with a prediction made by Lazarian \&
Pogosyan (2000) that the slope of a power spectrum should be steeper if
the emission comes from a relatively thick line of sight, meaning one with
a spatial thickness much larger than the transverse wavelength. Lazarian
et al. predict a slope of $-11/3$ in thick regions and $-8/3$ in thin
regions for {\it two} dimensional power spectra. These slopes increase
by 1 to $-8/3$ and $-5/3$ for
{\it one}-dimensional PS, which is appropriate for the present study.
In fact, the slopes in Figures \ref{fig:ps} and \ref{fig:psb}
equal these two values,
with a steepening from $-5/3$ at low spatial frequency to $-8/3$ at high
frequency.  This implies that the shortest wavelengths studied here are
significantly smaller than the line-of-sight thickness of the HI layer
in the Large Magellanic Cloud. The larger wavelengths where the PS is
shallow should be larger than the disk thickness. If this explanation
for the PS steepening is correct, then we are measuring for the first
time the line-of-sight depth of a gas layer in a galaxy. 

The steepening transition for the power spectra of the LMC occurs at a
wavelength of 80-100 pc. This is smaller than the expected thickness of
a dwarf galaxy but perhaps not unreasonable for the LMC.  Dwarf galaxy
thickness are several hundred parsecs (Hodge \& Hitchcock 1966; van
den Bergh 1988; Staveley-Smith et al. 1992; Carignan \& Purton 1998).
For the LMC, Wang \& Helfand (1991) estimate on the basis of a supershell
model for LMC2 that the thicknesses of the cool HI cloud layer and the
warm HI diffuse layer are 0.7 times the corresponding thicknesses in the
Milky Way.  This makes the cool cloud scale height 95 pc and the warm
diffuse scale height 350 pc. The full disk thicknesses would be twice
these values.  Kim et al. (1999) estimate an HI scale height of $\sim180$
pc for the LMC using dynamical parameters that have been determined for
the HI disk.  This derived scale-height is consistent with an apparent
distinction between giant shells and supergiant shells on a plot of
velocity versus radius (Kim et al. 1999).

Our HI observations come from a mixture of cool and warm HI blended into
a single spectral line, so a direct comparison with these thicknesses is
not possible at this time.  Model power spectra from layers with finite
thicknesses are presented in Section \ref{sect:models}.  They suggest
that most of the HI {\it structure} arises in a thin layer, presumably
from the cool component of the HI, while the thicker, warmer component
of the HI contributes little to these PS.  This result is consistent
with the observations of dwarf galaxies DDO 69 and Sag DIG by Young \&
Lo (1996, 1997), who found a low velocity dispersion (3.5 km s$^{-1}$)
from presumably cool HI in the central or clumpy parts of the galaxies,
and a high velocity dispersion (9 km s$^{-1}$) from presumably warm
HI elsewhere.  Most of the ISM structure is likely to be in the cool
phase because this compresses most easily.  If this cool phase has a
low cloud-to-cloud velocity dispersion, then it should lie close to a
galaxy's midplane.

There are several other ways to check if PS steepening is sensitive to
the disk thickness. First, the PS of whole-galaxy molecular maps should
have a steepening transition at a higher spatial frequency than the PS for
the HI because the molecular layer is probably thinner than the HI layer,
as observed for the Milky Way. Second, the wavelength at the steepening
transition should increase in the outer part of a face-on spiral galaxy
where the outer HI layer flares.  Similarly, the wavelength at PS
steepening should decrease for inner galaxy disks if the scale height
there is smaller as a result of the higher disk surface gravity. Higher
resolution and lower noise observations of the LMC should help clarify
this point as well, because they will extend the steep portion of the
PS further to higher frequencies.

The relatively steep slope of the power spectrum for the Milky Way
emission in Figure \ref{fig:foreground} 
implies that the emitting region
is physically thick there too, compared to the transverse wavelength.
For an average Milky Way distance of $\sim200 $ pc on the
line of sight, the longest transverse
wavelength is $\sim100$ pixels, or $\sim7$ pc.  Thus the emission
contributing to each channel has to have a depth on the line of sight
that is larger than this.

A different explanation for the steepening of the LMC power spectrum
at high spatial frequency is that the smallest clouds become optically
thick and appear smooth. In this case, the power spectrum for the total
column density could continue to decrease as a power law out to higher
frequencies than measured here, but the substructure below $\sim100$
pc may become washed out by the opacity effects, and cause the power
spectrum to drop faster.

\subsection{Spatial Variations in the Power Spectra as Evidence for
Localized Energy Input}
\label{sect:en}

\begin{figure}
\vspace{3.in}
\includegraphics{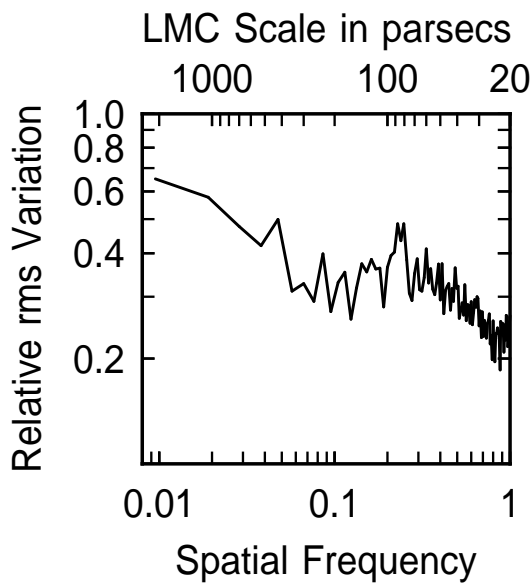}
\caption{Relative rms variations in the power spectra among all nine
subfields shown in figures 1 and 2. The relative rms variation is
largest on largest scales, to the left in the diagram, presumably
because turbulent energy is put into the interstellar medium
on these scales. }
\label{fig:variations}
\end{figure}

The variations of the PS from subfield to subfield are not noticeable in
Figure \ref{fig:ps} 
for scales less than $\sim200$ pc.  Larger scales
have obvious variations.  To study the variations better, Figure \ref{fig:variations} 
plots the relative rms deviations between subfields of the peak temperature
map,
given as the rms between all of the nine power spectra, normalized to the average power
spectrum. The relative rms variation
decreases slightly with wavenumber, from $\sim0.4$ to $\sim0.25$,
suggesting that field-to-field variations are slightly larger for longer
wavelengths. It is still relatively high at all wavelengths, though, as
if no particular wavelength dominates the field variations.  

We experimented with these rms variation by adding different amounts
of noise to the HI peak temperature map.  Random noise added to each
pixel caused the power spectrum to curl upward at high frequency,
with a greater curl for higher-amplitude noise.  This curl reflects the
increasing power at high frequency that comes directly from the noise.
However, the power spectrum for each of the 3x3 subfields curled up by
about the same amount, and this caused the relative rms variation from
subfield to subfield to decrease with increasing noise.  We conclude from
this experiment that noise in the data is not seriously affecting the rms
variations of the power spectrum from subfield to subfield.  In addition,
the average noise level in the observations is typically less than 10\%
of the peak temperature for most regions. This is too small to account for
the spatial variations in the power spectra.  Thus, these variations are most
likely the result of random positionings for real features in the maps.

The gradual decrease in subfield variations with wavenumber suggests
that energy is put in at long wavelengths to varying degrees in different
fields.  This is consistent with the haphazard placement of giant shells
throughout the LMC.  The similarity of the PS for all of the fields
at shorter wavelengths implies either that additional energy is put in
uniformly over a wide range of scales, or that the large-scale energy
cascades quickly and uniformly to smaller scales in each neighborhood.

\subsection{Power Spectra in the Azimuthal Direction}
\label{sect:az}

\begin{figure}
\vspace{6.in}
\includegraphics{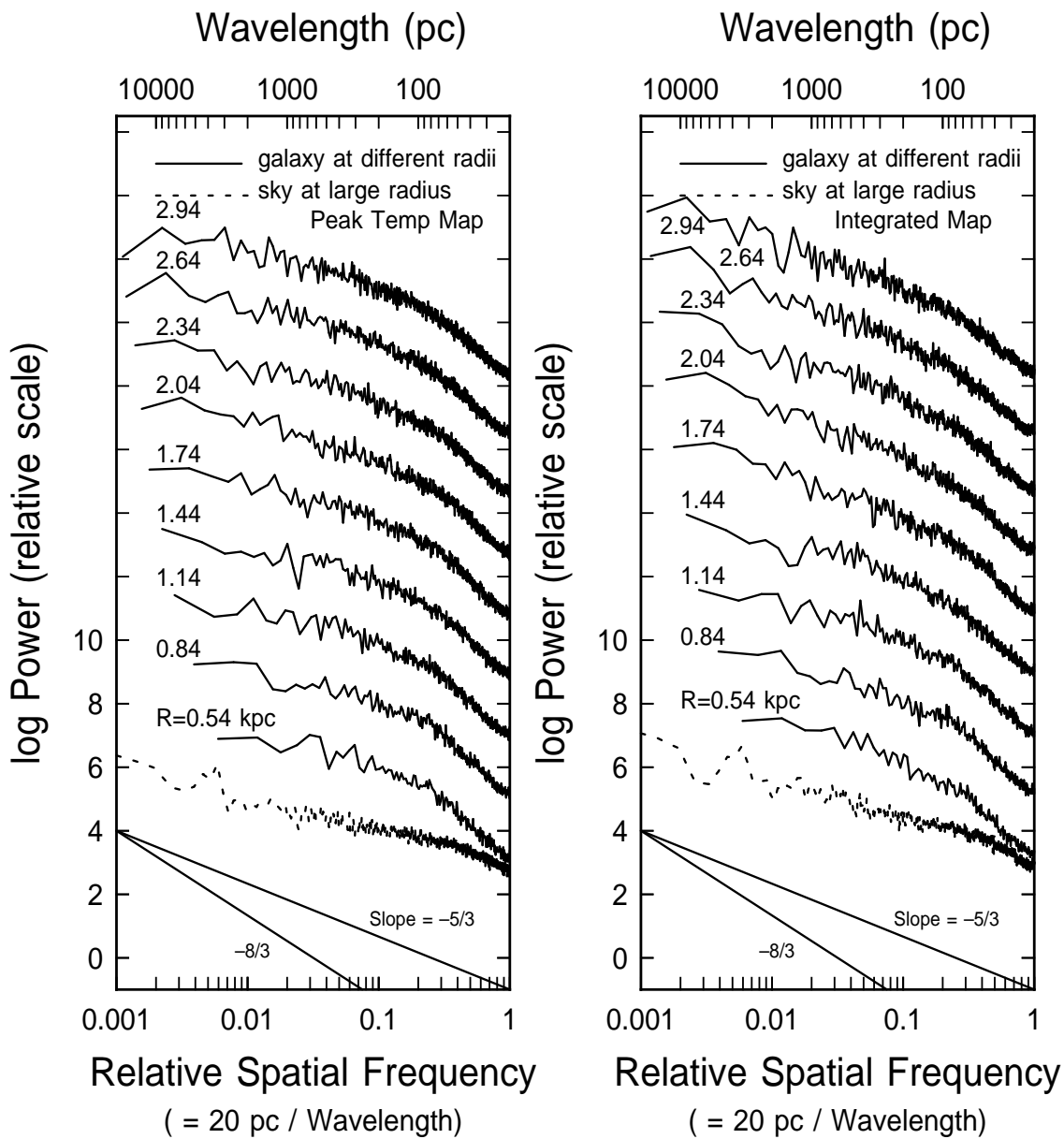}
\caption{Power spectra of the HI emission from the LMC
taken in deprojected azimuthal scans around the center
of the galaxy to remove the possible effects of the radial
disk gradient. The panel on the left is for the peak temperature
map, and the panel on the right is for the integrated map.  
Radii in kpc for each azimuthal scan are given next to the
corresponding curve.   The dashed line is from a sky region at 9.3 kpc,
shown on the same scale as the azimuthal power spectrum for
0.54 kpc.
The slope of the one-dimensional azimuthal power spectrum is
$\sim-5/3$ at small wavenumbers and $\sim-8/3$ at
large wavenumbers, the same as it is for the East-West and North-South
scans, indicating again some possible sensitivity to the line-of-sight
thickness of the HI layer in the LMC.  The wavelength of this
steepening point increases slightly with galactocentric radius, as
if the disk thickness increases with radius too. }
\label{fig:azimuth}
\end{figure}

Radial disk gradients and the disk edge
in the LMC may affect the PS at large scales, so we
removed any gradients by taking Fourier transforms of intensity scans
in the deprojected azimuthal direction.  Figure \ref{fig:azimuth} 
shows the average PS
of azimuthal scans from discrete radial intervals. Radius is
measured on the major axis, and the deprojection assumes
an inclination angle of $22^\circ$ and a position angle
of the line of nodes equal to $150^\circ$ from vertical in the map. 
Each PS is the average
of many PS, one for each pixel increment in radius within a range of radii
centered on the value indicated for each line in the figure. The intensity
at each azimuthal and radial position was determined by interpolation from
the Cartesian grid of intensities in the original map.  The left-hand
panel in the figure is for the peak temperature map, and the right-hand
panel is for the velocity-integrated map.  The PS for an azimuthal scan
at very large radius, 9.3 kpc, which is presumably in the sky region,
is shown as a dotted line at the bottom of each panel.  The abscissa
is normalized as in Figures \ref{fig:ps} and \ref{fig:psb}: 
single pixel fluctuations are
plotted on the right edge, where the relative spatial frequency equals
1, and longer wavelengths relative to this are plotted toward the left. The
longest wavelength increases with galactocentric distance as the physical
length of the circumference increases.  The radial intervals were chosen
so that the number of points in the azimuthal scans were permissible
ranges for FFT's in the IBM ESSL Fortran library.

The PS of the azimuthal scans in Figure \ref{fig:azimuth} 
show the
same steepening at large frequency as the PS of the horizontal or
vertical scans in Figures \ref{fig:ps} and \ref{fig:psb}. 
Each PS looks about the same,
aside from seemingly random fluctuations at a level of about $\pm0.4$
dex at low wavenumbers, decreasing to $\pm0.2$ dex at large wavenumbers.
This decrease is similar to that shown in Figure \ref{fig:variations}
and may result from enhanced energy input on large scales. 

There is a slight change in the frequency of the break point
as the radius of the azimuthal scan increases, from perhaps
80 pc wavelength at small radius to 150 pc wavelength at
large radius. This increase could be the result of
an increase in disk thickness with radius, as commonly
observed in galaxies. 

\subsection{Two-Dimensional Power Spectrum and Delta Variance}
\label{sect:2dps}

\begin{figure}
\vspace{3.in}
\includegraphics{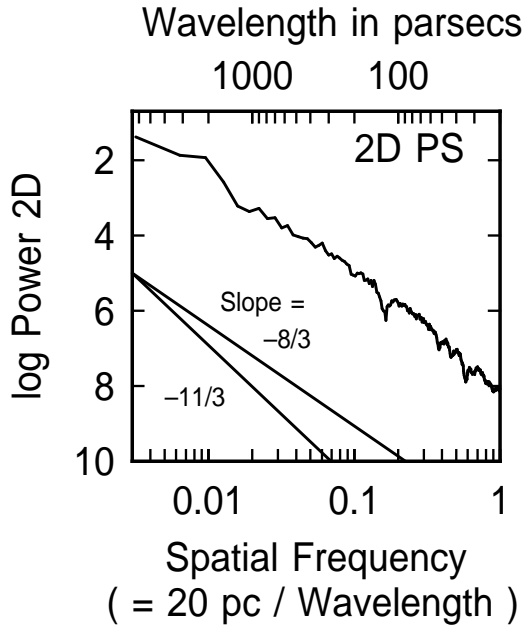}
\caption{Two-dimensional power spectrum
of the LMC integrated emission.} 
\label{fig:2dps}
\end{figure}

Figure \ref{fig:2dps} 
shows the two-dimensional power spectrum of the central
$630\times630$ pixel region ($\sim6.3\times6.3$ kpc)
of the integrated HI map of the LMC.
The plotted quantity is the sum of the squares of the real and the imaginary
parts of the 2D FFT made with the IBM ESSL subroutine package
SRCFT2, averaged over each 
2D wavenumber, $k=\left(k_x^2+k_y^2\right)^{1/2}$. 
The result is a power law with a slightly steepening slope, as indicated
by two lines.  The overall power-law behavior is similar to that
found by Stanimirovic et al. (1999)
for the 2D power spectrum of the SMC, but the slight steepening here was not
found there and is similar to that in the 1D power spectrum. 
The slope of the 2D power spectrum exceeds the average slope of the 1D power
spectrum by about 1, as expected for the higher dimension.

\begin{figure}
\vspace{3.in}
\includegraphics{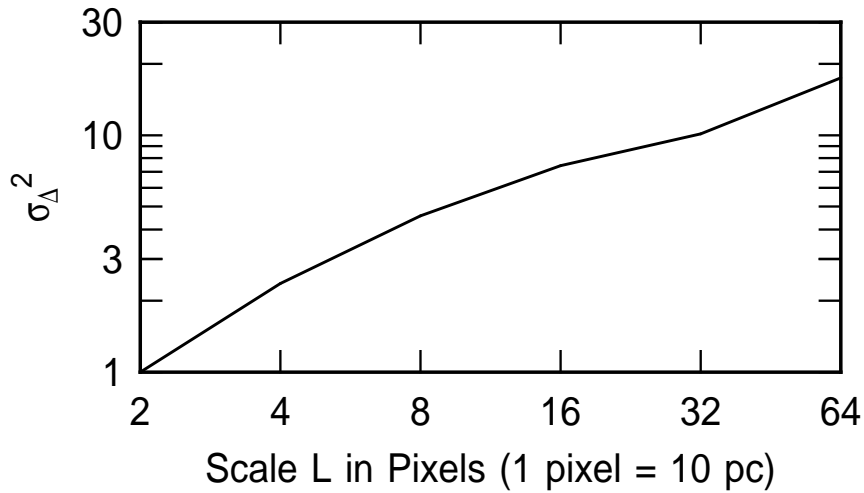}
\caption{Two-dimensional Delta variance 
of the LMC integrated emission.} 
\label{fig:2ddv}
\end{figure}

The two-dimensional Delta variance discussed by St\"utzki et al. (1998)
and Zielinsky \& St\"utzki (1999)
is shown in Figure \ref{fig:2ddv}.  This quantity is given by
\begin{equation}
\sigma_\Delta^2(L)=\int_{3L/2}^{630-3L/2} dx_0 \int_{3L/2}^{630-3L/2} dy_0 \int_{0}^{3L/2} dr
\left(\left[I(x_1,y_1)-<I> \right]
\* \bigodot \left[|r\{x_0,y_0\}-r\{x_1,y_1\}|\right]\right)^2
\label{eq:dv}
\end{equation}  
for function
\begin{equation}
\bigodot(r)=\pi \left(L/2\right)^{-2}\times\left\{1 \;\;
{\rm for}\;\;r<L/2\;\;,\;\; {\rm and} -0.125 \;\;{\rm for}\; L/2<r<3L/2\right\}.
\end{equation}
In this expression, the average intensity in the map is denoted by
$<I>$; $r(x_0,y_0)$ is the position of point
$(x_0,y_0)$ on the map, and
$r(x_1,y_1)$ is the position of point
$(x_1,y_1)$ on a small region around $(x_0,y_0)$; the distance between
these two points is 
$|r\{x_0,y_0\}-r\{x_1,y_1\}|$.

For the Delta variance, the length $L$ is measured in pixels
in powers of 2, from 2 to 64 pixels. Thus 
the $L=2$ case is like the integral over
an unsharp mask made by taking the difference between an 
average over a circle 1 pixel in radius, and the
average over the non-overlapping parts of a 
circle 3 pixels in radius; the $L=64$ case is about as big as we can get,
coming from  the difference between 
a radius-32 circle and the non-overlapping radius-96 circle.
The part of the whole map that is used for this integral, i.e., from
$3L/2$ to $630-3L/2$ pixels, decreases with $L$ in order to avoid 
spillover when the larger circle is subtracted. 

The Delta variance is expected to be a power law if the power spectrum
is a power law. The slope of the Delta variance on a log-log plot versus $L$
should be smaller than that
of the power law on a log-log plot versus $1/k$ by exactly 2.
This is about right in the figures here. The slope of the power law on a log-log
plot versus $1/k$ is 3.16, whereas the slope of the Delta variance on a log-log
plot versus $L$ is $\sim1$.

\begin{figure}
\vspace{6.in}
\includegraphics{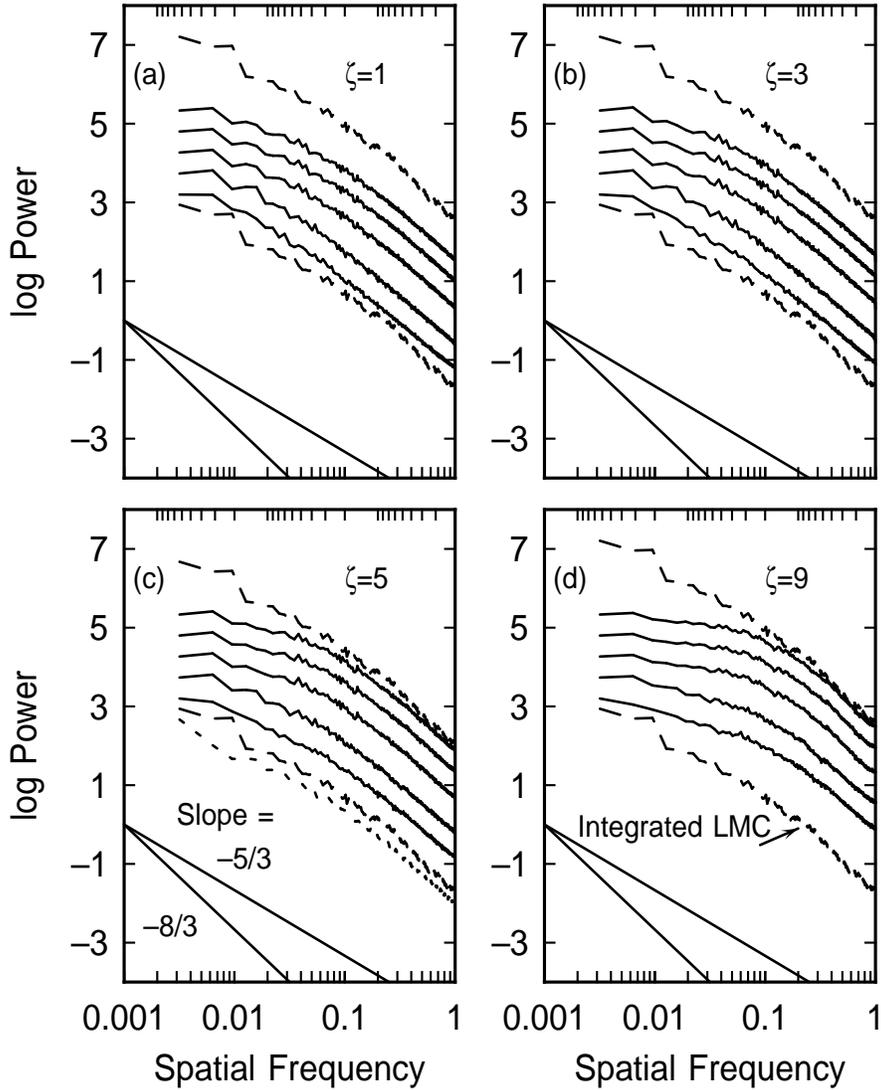}
\caption{Power spectra of the model fractal ISMs compared with the
observed power spectrum from the velocity-integrated map. The dotted
line in the lower right panel is for the model with an exponential disk.
It resembles the preferred model with the same parameters, 
$(\zeta,H)=(5,2)$, which is the
second line up from the dashed line, at all but the lowest frequencies.  }
\label{fig:fractalps}
\end{figure}

One advantage of the Delta variance over the power spectrum
is that the delta variance can avoid the image edges, while the
power spectrum assumes the image repeats itself infinitely
in a periodic fashion. This difference can be important for an image with sharp
edges, but for the case of a whole galaxy, where the edge tapers off gradually, 
the difference should not be physically significant. Our method to avoid the edges
with the Delta variance contains an error of its own in that we consider
different sub-parts of the whole galaxy for different scales $L$. 
Because the delta variance comes out to an approximate power law with
the right slope compared to the power spectrum slope, the edge effects
are not critical.  The dotted line in figure \ref{fig:fractalps}
below demonstrates this point in a different way.

\subsection{Filaments in the LMC}
\label{sect:fi}

A qualitative aspect of the HI distribution in the LMC that
cannot be obtained from power spectra is its filamentary nature. 
This is evident from the emission maps in Figure \ref{fig:images},
and is a well-known property of HI in the Milky Way
(Kulkarni \& Heiles 1987).

This filamentary structure becomes obvious if we compare
positive and negative images of the HI emission
in Figure \ref{fig:filaments}.
The lower panel shows the HI emission from a central region of
the LMC as a bright white color on a blue background. The top part
of the figure shows the HI emission as a blue color on a white background. 
The two parts of the figure look qualitatively different
because the HI emission is filamentary and the regions between the
HI emission, which are the intercloud medium, are more
globular. That is, the white features in the bottom figure are
filamentary, while the white features in the top figure are
globular; this is true even though these are exactly the
same figure with only the colors reversed.  A similar plot
(not shown) of a purely fractal cloud looks qualitatively
the same in each color scheme.

The emission regions are also more sharply peaked than the valleys
between them. Figure \ref{fig:lmc_line} 
shows a horizontal intensity 
trace 400 pixels long that comes from the middle of the lower figure in
Figure \ref{fig:filaments}.  The trace is shown in both linear
and logarithmic plots.  The linear plot has sharp maxima 
and shallow minimum, whereas the logarithmic plot has
more uniform sharpnesses for the maxima and minima.

\begin{figure}
\vspace{6.in}
\includegraphics{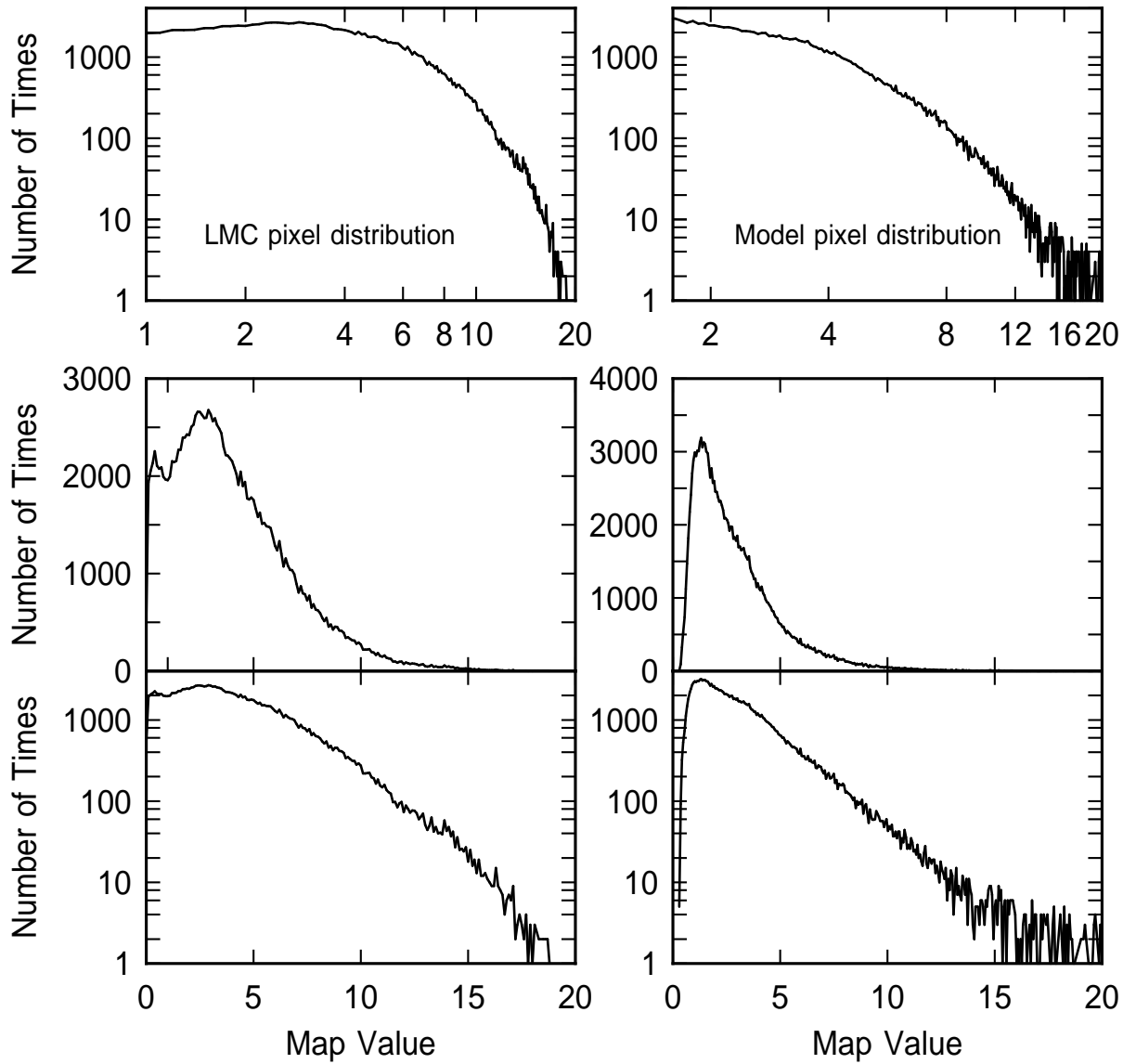}
\caption{(Left:) Histograms of the pixel values in a $400\times400$ pixel
field centered on the region shown in figure 10. 
The diagrams are plotted in log-log, linear-linear, and log-linear
coordinates.
(Right:) Histograms of the pixel distribution in the fractal model
shown on the top of figure 15.
}
\label{fig:his}
\end{figure}

The distribution function for the pixel-by-pixel values of intensity in
a $400\times400$ pixel region centered on the LMC field shown in figure 
\ref{fig:filaments} 
is plotted on the left in Figure \ref{fig:his} 
using
log-linear, linear-linear, 
and log-log coordinates.  The bottom panel shows that the
pixel values have an approximately Gaussian emission distribution. 
The curves on the right are from a model and will be discussed Sect. \ref{sect:models}.

\subsection{Unsharp Masks of the HI Emission}
\label{sect:ga}

Another aspect of HI gas structure is the possibility that the
morphology is different on different scales. The large scale seems
to be dominated by a few giant shells, whereas the small scale
seems filamentary
in some images. We can check this impression by plotting unsharp masks
of the data with different scales of smoothing, and then comparing
these to enlarged versions of the highest resolution mask, enlarged
to give the same number of pixels as the unsharp masks. 
If the enlarged version of a small region shows the same
character of structure as an unsharp mask
of a large region, blurred to the same number of pixels, then
the structure is qualitatively self-similar
on the two different scales. 

Figure \ref{fig:ga1} 
shows unsharp masks of the whole LMC, using
a $630\times630$ pixel field of view. 
The images were made as follows: first, each pixel of 
the integrated LMC map
was replaced by the average
of the surrounding $N_1\times N_1$ pixels. This made a blurred
map. Then another blurred map was made from averages over
$N_2\times N_2$ pixels.  The final unsharp mask image is
the difference between these two blurred maps.

In Figure \ref{fig:ga1}, the panel in the top left was
made from the difference between $3\times3$ and $11\times11$ maps;
the top right is the difference between
$11\times11$ and $41\times41$ maps, the bottom left is the
difference between $21\times21$ and $81\times81$ maps, and
the bottom right is the difference between $41\times41$
and $161\times161$ maps.  Thus the images highlight structures
on scales of 3, 11, 21, and 41 pixels. Recall that each pixel
is about 10 pc for the LMC.  There is no significant error from
overflow outside the image in the figure because there is high-quality
data from sky regions
outside the galaxy for all of the pixels used in these unsharp masks
(e.g., Fig. \ref{fig:sky}).

Figure \ref{fig:ga2} repeats the top left image from Figure 
\ref{fig:ga1}
and also shows enlarged images of this map from the same
position. The enlargements are by factors of $11/3$, $21/3$, and $41/3$,
which means that the enlarged structures from the highest-resolution
unsharp masked image have the same number of pixels as the smoothed
structures in the three unsharp masked images.  The enlarged images
have the same character of structure as the corresponding unsharp masked
images, although not the same total range of intensities.  The overall
structure may be characterized as having holes with filamentary emission
regions surrounding them.  This implies that the geometry of the structure
is qualitatively self-similar on the range of scales covered, which
corresponds to 30 pc for the $3\times3$ pixel unsharp mask image to 410
pc for the $41\times41$ image.  The self-similarity could persist up to
even larger scales, but boundary effects make it hard to get a smoothed
background on these larger scales for subtraction in the mask.

The range of intensities for the unsharp masked and enlarged images
is different.  There is no analogy to the extremely
bright region south of 30 Dor in the enlarged region. 

\section{Fractal Emission Models}
\label{sect:models}

\subsection{Model Setup}

The wide range of scales that is observed for interstellar structures
has led to a fractal model, starting with early observations of dust
clouds (Beech 1987; Bazell \& D\'esert 1988; Scalo 1990), molecular
clouds (Dickman, Horvath, \& Margulis 1990; Falgarone, Phillips,
\& Walker 1991), local atomic clouds (Vogelaar \& Wakker 1994), and
more recently, the ionized intercloud medium (Berkhuijsen 1999) and
whole galaxies (Stanimirovic et al. 1999; Westpfahl et al.  1999).
Computer simulations of supersonic turbulence also have a scale-free
or fractal quality (Elmegreen 1999; Rosolowsky et al. 1999; MacLow \&
Ossenkopf 2000; Pichardo et al. 2000).

Because we observe a scale-free quality to the HI emission from the 
Large Magellanic Cloud, from both power spectra and unsharp masks, 
we would like to model the LMC with an idealistic
fractal that is reasonably close to what might be expected from
turbulence.  By varying the line of sight depth and other
properties of the model and fitting the resultant map to the LMC 
observations, we might be able to measure specific properties of the
observed interstellar gas.  

With this in mind, 
we made three-dimensional models of a fractal gas using a technique
pioneered by Voss (1988). In this method, random complex-number
noise in 3D wavenumber space is multiplied by a power law, and then
the inverse Fourier transform is taken to give a fractal distribution
in real 3D space.  To reproduce something like the density structure
that is expected from turbulence, we did two special things. First,
the power law multiplier in wavenumber space was taken to be $k^{-5/3}$
for 3D wavenumber $k$.  Second, after the 3D inverse Fourier transform
was taken, the resulting pixel values, X, were exponentiated, making a 3D
cube of values given by $\exp(\zeta X)$ instead of just $X$.  This second
step produces a 3D fractal with a near-power law power spectrum, and also with
a probability distribution function of density ($=$ final pixel values)
that is log-normal. Such a log-normal pdf for density has been obtained from
3D numerical simulations of MHD turbulence (Vazquez-Semadeni 1994;
Scalo et al. 1998; Nordlund \& Padoan 1999). 

We infer from the results in Nordlund \& Padoan (1999)
that the coefficient in the
exponent, $\zeta$, should be related to the Mach number, $M$, as: 
\begin{equation}
\zeta = {{\left(\ln\left[1+0.5M^2\right]\right)^{0.5}}\over
{X_0}}.
\label{eq:xi}
\end{equation}  
This relation arises as follows: the distribution function of
pixel values that return from the inverse Fourier transform 
of the noise is found to be
approximately a Gaussian, $\exp\left(-0.5\left[X/X_0\right]^2\right)$, 
with a dispersion in the cases we ran equal to $X_0=1.27$.
We use this fact to associate $X$ with the log of the
density, writing for probability functions:
\begin{equation}
P_{\ln\rho}\left(\ln\rho\right)d\ln\rho = P_{X}\left(X\right)dX=
e^{-0.5\left(X/X_0\right)^2}dX .
\end{equation}
With $\zeta X= \ln \rho$, this becomes
\begin{equation}
P_{\ln\rho}\left(\ln\rho\right)=
e^{-0.5\left(\ln\rho/\left[\zeta X_0\right]\right)^2}{{dX}\over
{d\ln\rho}}=
e^{-0.5\left(\ln\rho/\sigma\right)^2}{1\over{\zeta}},
\end{equation} 
where $\sigma=\zeta X_0$ is the dispersion of the pdf for
density. This dispersion has been found by 
Nordlund \& Padoan (1999) to be
$\sigma = \left(\ln\left[1+0.5M^2\right]\right)^{0.5},$
from which equation \ref{eq:xi} follows. 

Numerical simulations suggest that the density distribution function has
a log-normal form only for isothermal turbulence, and that a power law
function might be better for other cases (Scalo et al. 1998).  One of
the reasons for this is that a power law has much more structure at high
densities than a log-normal, and this is the result of the presence
of dense cool clouds in a non-isothermal model.  With only one phase
of matter in the isothermal case, the density cannot have such large
variations at modest Mach numbers.  The difference between these two
cases has not been recognized yet from real power spectrum data, although
we know that the ISM in the LMC does have two phases of HI from other
studies (Marx-Zimmer et al. 2000).  In what follows, we demonstrate
that the structure of the ISM in the LMC requires more than isothermal
turbulence. We do this by first obtaining an absurd result when
the above model based on isothermal turbulent structure is assumed.
We then modify the model to simulate two phases and obtain a nice fit
to the observations.

Various values of the Mach-number parameter $\zeta$ were used in different
models, along with various line-of-sight depths to an integral
over the fractal density distribution, in an attempt to fit both the
observed power spectrum of the LMC and the amplitude and character of the
intensity trace shown in Figure \ref{fig:lmc_line}.  We used $\zeta=1,$
3, 5, and 9 to simulate different Mach numbers and therefore different
compressions from turbulence.  For the model intensity, we "observed"
the density distribution by integrating along one dimension of the
model fractal, taken to represent distance on the line of sight. To
simulate the thin disk of a galaxy, the 3D fractal was multiplied by a
Gaussian centered on the middle pixel of the line of sight and having
a scale height $H$. This represents a Gaussian ISM in a face-on galaxy,
and the integral over this dimension represents the velocity-integrated
spatial map.  To compare with the observations of the LMC, we took 1D
power spectra of the model map, averaged over all horizontal strips.

Several maps were made with various $H$. All had total transverse
dimensions of $720\times720$ pixels, of which only the central
$630\times630 $ pixels were used for the power spectrum analysis, to avoid
possible edge effects.  This size is comparable to the total field size of
the LMC map used for the power spectrum analysis in section \ref{sect:ps}
(the LMC field size was $630\times600$, with the 630-pixel dimension in
the direction of the one-dimensional power spectrum).

The depth of the 3D density distribution was always 60 pixels, which
is the largest size that could fit in the memory of a single node of
our computer, considering the other dimensions were $720\times720$.
In one case, only the central pixel was taken for the sky map, and in
other cases, Gaussian multipliers for the fractal density distributions
on the line of sight were used with Gaussian dispersions of $H=0.125,$
0.5, 2, and 8 pixels.

A spatial map for the $\zeta=5,$ $H=2$ case is shown on the top of
figure \ref{fig:fractalmap}.  This case is chosen for display because
it is a good fit to the power spectrum given in the next sub-section.
Considering the pixel size in the LMC map, this $H=2$ case corresponds to
a scale height for the fractal structure of $\sim20$ pc. On the bottom of
figure \ref{fig:fractalmap} is the same map multiplied by an exponential
with four scale lengths out to the edge.  This is meant to simulate the
exponential disk in a real galaxy. The purpose of this exponential model
is to demonstrate that edge effects in the sharp-edged fractal models
do not cause the observed curvature and bends in the LMC power spectra
(see next sub-section).

\subsection{One dimensional Power Spectra and Intensity Traces of Models}

Figure \ref{fig:fractalps} shows the 1D power spectra of the models in
four panels representing the different $\zeta$ values.  In each panel,
the order from top to bottom of the plotted power spectra is an order of
increasing thickness on the line of sight: the single pixel result is at
the top, and the $H=$ 0.125, 0.5, 2, and 8 cases follow.  The dashed line
is the observed LMC power spectrum from the integrated map, which was also
shown as a dashed line in the middle plot of Figure \ref{fig:psb}.  It is
drawn twice in each panel to allow easy comparison with the model results.
The dotted line is the $(\zeta,H)=(5,2)$ exponential disk case shown at
the bottom of figure \ref{fig:fractalmap}.  Not all of the model PS fit
the observations. Those which do are the models with $(\zeta,H)=(3,2)$
and $(5,2)$.  The difference between the exponential disk case and the
sharp-edge case for the same $(\zeta,H)$ is apparent only at very low
frequency.  Thus the sharp edges in most of the fractal models do not
affect our interpretation of the curvature in the power spectra.

\begin{figure}
\vspace{6.in}
\includegraphics{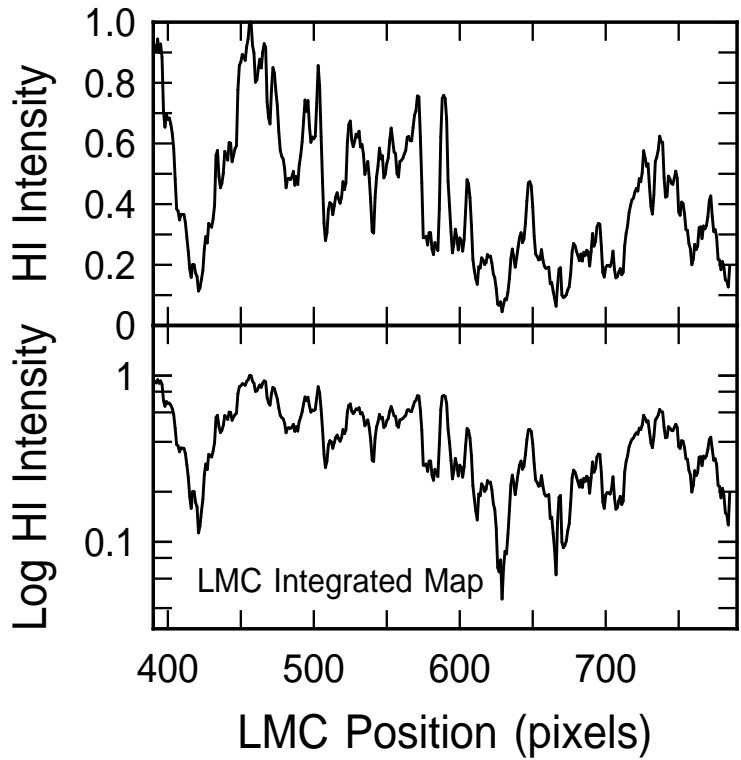}
\caption{An HI intensity strip from the central region of the LMC
field in figure 10, shown in linear and logarithmic
coordinates. The fluctuations around the mean are more symmetric
and random-looking
on the logarithmic plot, 
illustrating the peaked-nature of the emission regions, and the
broad nature of the intercloud regions. 
}
\label{fig:lmc_line}
\end{figure}

\begin{figure}
\vspace{6.in}
\includegraphics{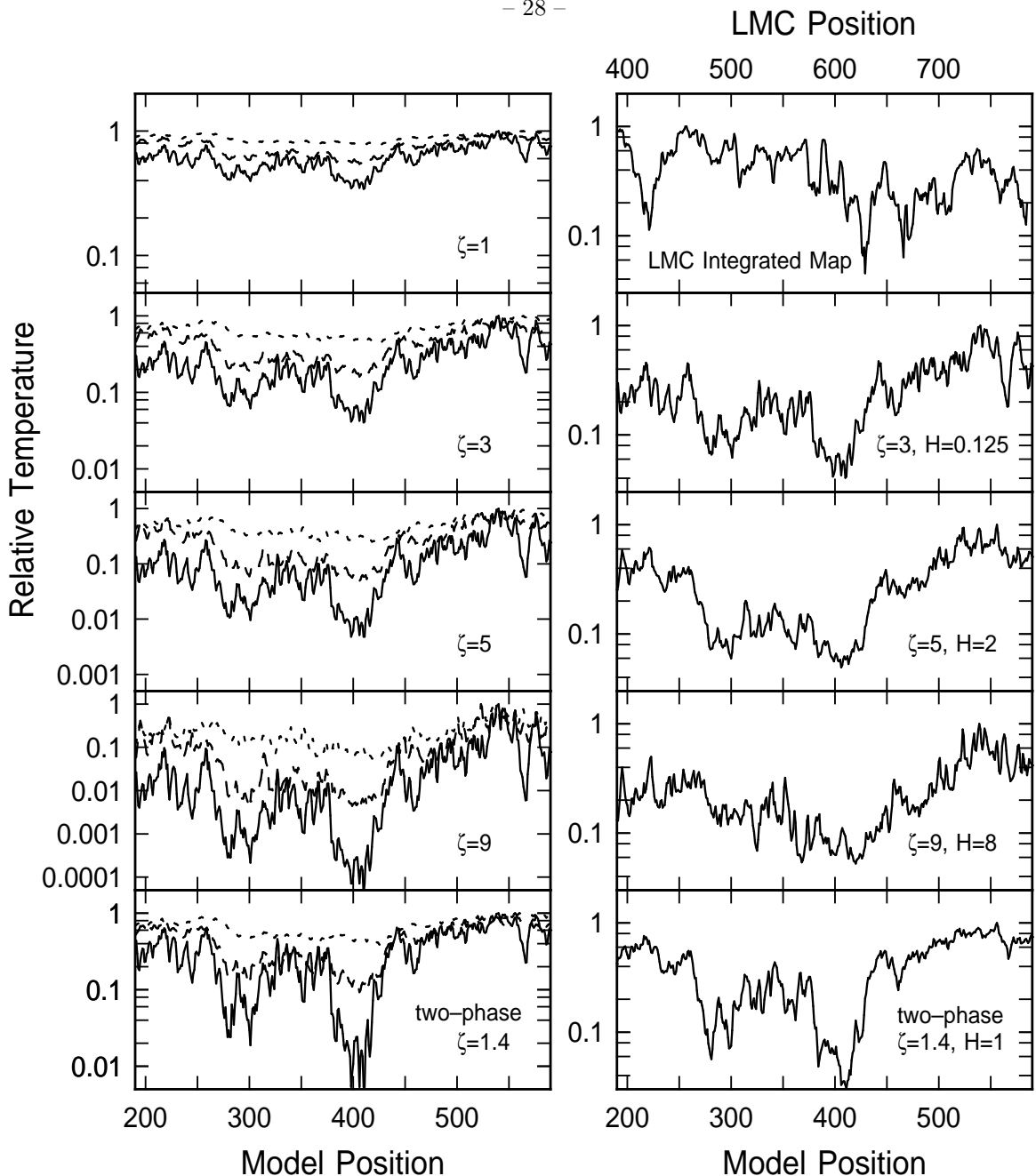}
\caption{Intensity traces from various fractal models
compared to the observed HI intensity trace (top right) from the
field in figure 10, also shown in figure 11. 
The dotted, dashed and solid lines on the left are for models with
different scale heights in the Gaussian representation of the
face-on galaxy: $H=0.125$, 2, and 8, respectively. 
The different panels are for different Mach number parameters, $\zeta$,
with larger $\zeta$ representing larger Mach numbers. 
Acceptable models have intensity fluctuations of a factor of $\sim10$,
and are shown on the right for each Mach number parameter, $\zeta$. 
The bottom panels are for the model with a simulated two-phase
medium. 
}
\label{fig:lines}
\end{figure}

Figure \ref{fig:lines} 
shows a horizontal trace of intensity from
the middle of each model map, made in the same way as in Figure 
\ref{fig:lmc_line}. The observations are reproduced in the top right of
this figure, and a range of models is on the left. Each panel on the left
shows a different $\zeta$ value, and each line shows a different $H$,
taken to be the $H=0.125$, 2 and 8 models.  The $H=0.125$ results have
the largest amplitudes and are represented by solid lines, while the
$H=8$ lines have the lowest amplitudes and are represented by dotted
lines. The amplitude of the variation in the intensity trace decreases
with increasing depth over the line of sight because the additional random
HI washes out the total emission on a deep line of sight.  The amplitude
variation also decreases with decreasing $\zeta$ because the density
fluctuations are smaller when the Mach number is smaller.

The right hand side of Figure \ref{fig:lines} 
shows again the $H$
values from the left that reproduce best the observed amplitude of the
HI intensity variation. The amplitude variation is about a factor of
ten. All of the curves on the right are to the same scale as the LMC
intensity strip in both pixel resolution and amplitude.  All of the
curves also use exactly the same strip in the model fractal, so the
variation among them is entirely from the variations in $\zeta$ and $H$,
and not from a variation in the noise distribution in wavenumber space.

The top four left-hand panels in Figure 
\ref{fig:lines} 
are for this standard
fractal ISM model. The bottom panels are for a model with a simulated
two-phase HI structure, as discussed below.  The best fit models in
the standard cases have $(\zeta,H)=(3,0.125)$, $(5,2)$, and $(9,8)$,
as shown on the right. An
increase in $\zeta$ gives more density contrast and an increase in $H$
gives less column density contrast, so $\zeta$ and $H$ increase together
to give the same overall column density contrast.

The only standard fractal model that fits both the 1D power spectrum of
the observations and the amplitude of the intensity variations in the
LMC data is the model with $(\zeta,H)=(5,2)$.  This is the model shown
previously on the top of Figure \ref{fig:fractalmap}.  Generally this
model is a good fit to the LMC, in terms of power spectrum and intensity
fluctuations. The character of the intensity strip is a good match too,
with similar excursions on both large and small scales.  This latter
point is shown in Figure \ref{fig:his}, where the distribution functions
for pixel values in the LMC data (left) and the fractal model (right)
with $(\zeta,H)=(5,2)$ are shown in various coordinate systems.
The approximately Gaussian character of the pixel distribution of the
observed HI intensity is also present in the model intensity distribution.

The standard fractal model differs from the real LMC in several ways,
however. The model lacks the obvious filaments, holes, and shells
that are seen by eye in the LMC, and it lacks the sharp transitions
from bright to dim that are seen in the LMC intensity strip of Figure 
\ref{fig:lmc_line}. These sharp transitions may be phase transitions
in the HI, or they could be shock fronts. They contribute only a small
fraction of the total LMC HI luminosity, and so do not influence the
power spectrum much. They are probably related to the mechanisms of
cloud formation however, and are therefore important physically.

\subsection{Model with Two Thermal Phases}

There is another difference between the model results and the observations
and that is the Mach number corresponding to the best-fit case,
with $\zeta=5$. According to equation \ref{eq:xi}, the Mach number
is extraordinarily large, $\sim10^9$. This is an absurd result and it
indicates that the model, even with a turbulence-quality in terms of
density distribution, does not contain all of the physics of the real
LMC. The real HI distribution has much larger column density variations
than can possibly come from homogeneous turbulence alone. Much of
the observed variation has to come from other causes, such as phase
transitions making low intensity levels all along certain lines of sight,
and sharply delineated shock fronts that do the same. A more reasonable
Mach number would have $\zeta\sim1$, but then the amplitude variation of
the intensity strip is too small by a factor of $\sim3$, according to
the curves in the top left of Figure \ref{fig:lines}. The remaining
factor of $\sim3$ in the observed intensity variation probably comes
from phase transitions in the HI gas and other density fluctuations
independent of pure turbulence compression.

\begin{figure}
\vspace{3.in}
\includegraphics{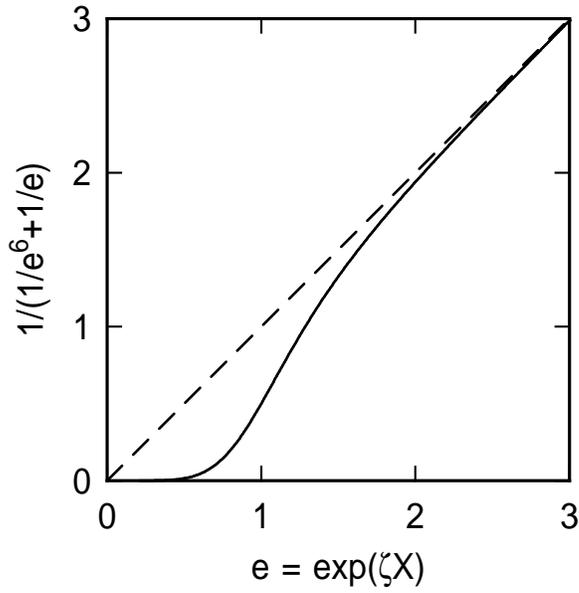}
\caption{Transformation of density in the original fractal model,
$\rho_0=\exp(\zeta X)$, into density in a simulated two-phase model, 
$\rho_{\rm two\;phase}$. Low values of the original density are
converted into very low values in the two-phase model, while
high values are not changed much.}
\label{fig:transform}
\end{figure}

We simulated the additional density structure from HI phase
transitions in a fractal ISM model with an effective low Mach
number by converting the pixel density values obtained from the
previous fractal model, $\rho_0=\exp(\zeta X)$, into a two-phase
density structure using a convenient transformation \begin{equation}
\rho_{\rm two\;phase}={{1}\over{\rho_0^{-6}+\rho_0}}.  \label{eq:cool}
\end{equation} This transformation is shown in Figure 
\ref{fig:transform}.
High pixel values, representing high-density clouds, are not changed
much, while low pixel values, representing the intercloud medium, are
reduced to much lower values.  The intensity traces from this two-phase
model are shown at the bottom of Figure \ref{fig:lines}; the $H=0.125$,
2 and 8 cases are on the left, as before, and the $H=1$ case is on the
right as a best fit to the observed amplitude variation. A low value of
$\zeta=1.4$ was assumed.

\begin{figure}
\vspace{3.in}
\includegraphics{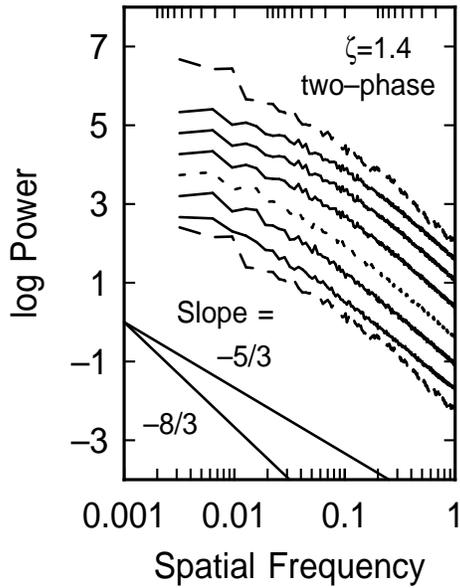}
\caption{Power spectra of the models with simulated two phases, as in 
figure 16.}
\label{fig:pscool}
\end{figure}

The 1D power spectra of the two-phase models are in Figure 
\ref{fig:pscool}.
The values of $H$ used previously in Figure 
\ref{fig:fractalps}
are used again here, as solid lines, but now the case with $H=1$
pixel is also shown as a dotted line because that is the best fit in
Figure \ref{fig:lines}.  A map of the column density structure for
the two-phase model with $(\zeta,H)=(1.4,1)$ looks essentially the
same as the map in Figure \ref{fig:fractalmap} 
and is not shown.

For the value of $\zeta=1.4$ in the two-phase model, the Mach number from
equation (\ref{eq:xi}) equals 9.5, which is a reasonable value for a
galaxy disk.  Higher values of the power in the denominator of equation
(\ref{eq:cool}), as well as smaller values of $H$,
give the same intensity variations with lower $\zeta$.

\subsection{Two-Dimensional Power Spectra and Delta Variances
of the Two-Phase Models}

\begin{figure}
\vspace{4.in}
\includegraphics{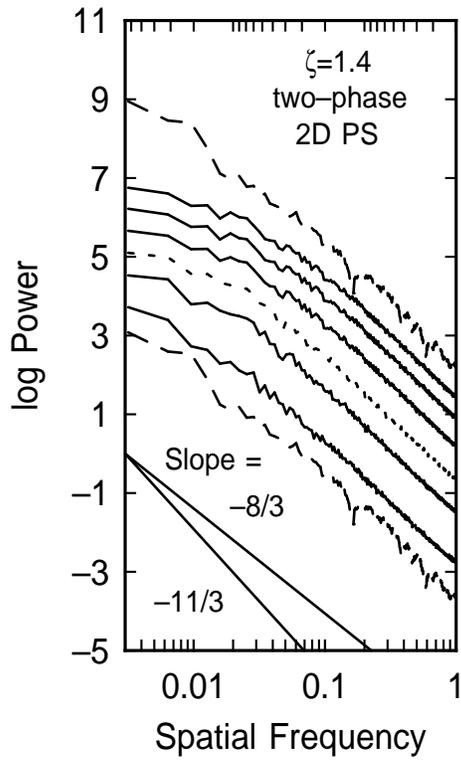}
\caption{Two-dimensional power 
spectra of the models with simulated two phases.}
\label{fig:tdpscool}
\end{figure}

Two-dimensional power spectra of the two-phase models are shown in Figure
\ref{fig:tdpscool}, compared to the 2D PS of the integrated map, which is
a dashed line plotted twice for clarity.  All the PS have been normalized
to have unit value at the lowest wavenumber, and they have been shifted
vertically in the plot for clarity.  The order of the model PS is, from
top to bottom: single channel integral, $H=0.125$, $0.5,$ 1, 2, and 5,
with $H=1$ shown as a short-dashed line, as in Figure \ref{fig:pscool}.
Again, the models with low $H$ have too shallow a power spectrum at
low frequency, and the models with large $H$ have slightly too steep a
power spectrum at low frequency. The $H=1$ pixel case (corresponding to
$\sim10$ pc) is about right.  The power law for the real LMC data depends
on several major asymmetries,
such as the enhanced brightness south of the 30 Dor region.  These are
not present in the fractal model, so the lowest few frequency points on
the dashed (LMC) line of Figure \ref{fig:tdpscool} are not expected to
fit the model well.  The rest of the model is reasonably good though.

\begin{figure}
\vspace{2.5in}
\includegraphics{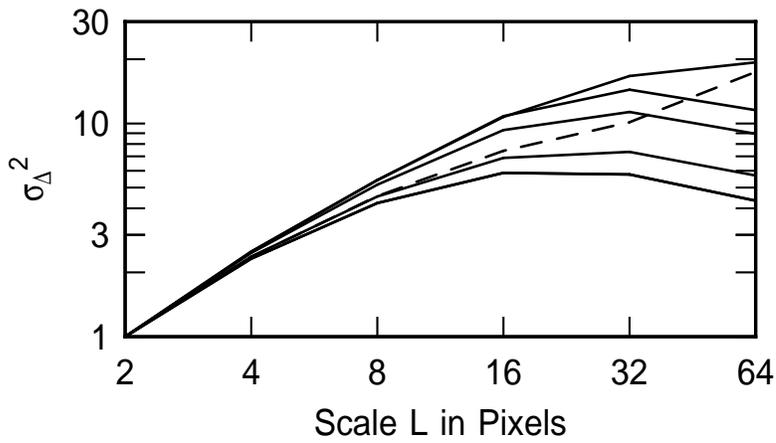}
\caption{Two-dimensional Delta variances
of the models with simulated two phases.}
\label{fig:2ddeltacool}
\end{figure}

The Delta variances of the two-phase models are shown as solid lines in
Figure \ref{fig:2ddeltacool}.  The dashed line is the Delta variance for
the LMC from Figure \ref{fig:2ddv}.  The line-of-sight depth increases
as the model curves go from top to bottom.  The top curve has a single
channel in depth, and another at essentially the same place has a Gaussian
width on the line-of-sight equal to 0.125 pixel.  The remaining have
Gaussian widths of 0.5, 1, 2, and 8 pixel, as in the previous figures.
The slope of the Delta variance at small scales is 2 different from the
slope of the power spectrum, as expected, and it is in good agreement
with the Delta variance for the observations. The model slopes fall for
larger scales, presumably because of the growing importance of a few large
features inside an ever-decreasing boundary (e.g. St\"utzki et al. 1998).

\section{Conclusions}

The structure of HI in the LMC shows the same scale-free character in
two-dimensional power spectra as
it does in the SMC (Stanimirovic et al. 1999), but it 
steepens at short wavelengths in the one and two-dimensional power spectra,
and in the azimuthal power spectra,
perhaps as a result of a transition to a relatively thick disk on
that scale. Such steepening has been predicted for such a transition
from thin to thick lines of sight (Lazarian \& Pogosyan 2000), but its
importance as a potential indicator of extragalactic disk thickness was
not previously recognized.

The power spectrum of the whole galaxy resembles the power spectra of
individual channel maps in the regions where the channel maps trace
the velocity centroids of the emission.  This implies that individual
channels contain essentially the same structure as the whole spectral
line at the spatial position where the centroid of the line has the same
velocity as the channel map.  Narrow velocity slices do not come from a
thinner region than the whole spectral line, for example, and the small
scale structure in the ISM is not limited to any particular part of the
line profile.  This conclusion is based only on our study of very large
regions and does not necessarily apply to all lines of sight. Expanding
shells should deviate from this trend, of course, because the small near-
and far-side caps should appear in different parts of the spectral line
than the large projected circumference.

The character of the HI structure was also investigated using positive and
negative images to compare the high and low emission regions, and using
unsharp masks with four different resolutions.  On all scales between
$\sim30$ and $\sim 400$ pc, the emission tends to be filamentary, and
the regions between the emission tend to be round. This is what might
be expected from a conglomerate of expansion sites, although the nature
of the high pressure that drives each expansion is not indicated. For
example, the high pressure could be stellar in origin, or it could come
from turbulent fluctuations.

Fractal models of a face-on galaxy with a log-normal probability
distribution function for local density and a Gaussian distribution of
density on the line of sight reproduce the observations fairly well. The
models get the power spectrum of the LMC, the peak-to-peak amplitude of
the emission variations, and the probability distribution function for
pixel values.  The models suggest that most of the spatial structure
in the HI is in a relatively thin midplane layer, perhaps $\sim50$
pc thick. This implies that any thicker HI layer that surrounds the
cloudy disk has relatively little structure; i.e., most of the
LMC clouds are in a cool component of the HI that has a relatively low
velocity dispersion and a low scale height.

The models include a log-normal distribution for local density in
order to simulate compressible isothermal turbulence.  The model can be tuned to
the observations by adjusting the Mach number of the turbulence, which
enters into the Gaussian dispersion of the log-normal, and the thickness
of the disk.  The best fits require unrealistically large Mach numbers,
so other models were made having additional HI structure from a simulated
phase transition between warm and cool HI components.  These models
could be tuned to fit the LMC data with a more realistic Mach number.

Positional fluctuations in the power spectra are slightly larger at larger
scales, suggesting that significant energy is put into the interstellar
medium on kiloparsec scales, probably in connection with the well-known
supershells. However, the positional fluctuations are still fairly
large on smaller scales, amounting to 0.2-0.3 dex, so there are either
additional energy sources on small  scales, or quick and local cascades
of supershell energy to smaller scales.  A power-law distribution of shell
sizes (Oey \& Clarke 1997) may provide the range of scales necessary.

The power spectra of HI emission at large scales in the LMC are
slightly shallower than the power spectra of foreground Milky Way
emission. Presumably the foreground emission is thick on the line of
sight, relative to the largest transverse wavelength, as is the short
wavelength emission from the LMC. The similarity of the power spectra
for thick emission is remarkable considering the factor of $\sim200$
difference in physical scales. 

We would like to acknowledge the encouragement and advice of Dr. Agris
Kalnjas at Mount Stromlo Observatory in Australia. Helpful comments on
the Delta variance by Drs. Ossenkopf and Mac Low, and on the power
spectra of turbulent gases by Dr. Lazarian, are also appreciated. This
work was supported by NSF Grant AST-9870112 to B.G.E.

\newpage

\begin{figure}

\caption{Four images of HI emission from 
the LMC, each 
with $3\times3$ subfields for separate power spectrum analyses. 
Top left: peak HI temperature;
top right: HI integrated over velocity; Lower left: channel map at
$254\pm0.8$ km s$^{-1}$; lower right: channel map at 
$302\pm0.8$ km s$^{-1}$. The giant HII region 30 Dor is in the
middle subfield on the left; the supershell LMC4 is on the white
line dividing the top left and middle subfields. (see jpg file for astroph)}
\label{fig:images}
\end{figure}

\newpage

\newpage

\begin{figure}
\caption{HI emission from the Northeast corner of the velocity-integrated
map, showing a box in which the power spectrum of the sky was calculated.
(see jpg file for astroph)}
\label{fig:sky}
\end{figure}

\newpage

\newpage

\newpage

\begin{figure}
\caption{Colorized emission maps of a central region of the LMC, showing
white HI emission on a blue background at the bottom of the figure, and
the reverse colors at the top. The difference in morphology of the white
regions in these two images illustrates the difference between the cloud
geometry and the intercloud geometry. The clouds are often filamentary,
while the intercloud medium is often globular. This difference suggests
the HI structure is made by expansion around intercloud centers, either
from turbulence or from point sources of energy. (see jpg image for astroph)
} \label{fig:filaments}
\end{figure}

\newpage

\newpage

\begin{figure}
\caption{Unsharp masks of the integrated HI emission from the whole
LMC, emphasizing features on scales of 30 pc, 110 pc, 210 pc, and 410 pc. 
(see jpg image for astroph) }
\label{fig:ga1}
\end{figure}

\begin{figure}
\caption{Expanded versions of the high resolution unsharp mask in the
upper left, which is reproduced from figure 13. 
The expansions give the other maps here the same pixel resolution as
the corresponding maps in figure 13.  The character of
the emission, with circular holes surrounded by filamentary 
emission, is approximately the 
same on all of the scales shown, which range from 30 pc to 410 pc.
(see jpg image for astroph) }
\label{fig:ga2}
\end{figure}

\begin{figure}
\caption{top: Column density map of a standard 
model fractal ISM measuring 630x630
pixels. The model has a Gaussian density distribution on the line
of sight with a dispersion of $H=2$ pixels to represent the
thin layer of a face-on galaxy. A turbulence parameter
representing Mach number has the value $\zeta=5$. The density in each
pixel of the volume fractal is given by $\exp(\zeta X)$ for
a fractal field $X$ generated from the inverse Fourier
transform of noise with a power-law cutoff. 
The two-phase model looks about the same as this.
bottom: The same model with an exponential taper 
having four scale lengths to the edge. (see jpg image for astroph) }
\label{fig:fractalmap}
\end{figure}

{\bf Figure Captions}

\newpage

\newpage

\newpage

\end{document}